\documentclass[aps,prb,twocolumn,10pt,superscriptaddress,showpacs,nofootinbib]{revtex4-1}
\usepackage{graphicx}
\usepackage{bpchem}
\usepackage{amssymb}
\usepackage{color}
\begin{document}

%\preprint{}

\title{Magnetocrystalline Anisotropy and the Magnetocaloric Effect in \BPChem{Fe\_{2}P}}


\author{L. Caron} 
\email[]{L.Caron@tudelft.nl}
\affiliation{Fundamental Aspects of Materials and Energy, Faculty of Applied Sciences, TUDelft\\ Mekelweg 15, 2629 JB Delft, The Netherlands}
\affiliation{Department of Engineering Sciences, Uppsala University, Box 534, SE-751 21 Uppsala, Sweden}
\author{M. Hudl}
\affiliation{Department of Engineering Sciences, Uppsala University, Box 534, SE-751 21 Uppsala, Sweden}
\author{ V. H\"{o}glin} 
\affiliation{Department of Materials Chemistry, Uppsala University, P.O. Box 538, SE-751 21 Uppsala, Sweden}
\author{N. H. Dung}
\affiliation{Fundamental Aspects of Materials and Energy, Faculty of Applied Sciences, TUDelft\\ Mekelweg 15, 2629 JB Delft, The Netherlands}
\author{C. P. Gomez}
\affiliation{Department of Materials Chemistry, Uppsala University, P.O. Box 538, SE-751 21 Uppsala, Sweden}
\author{M. Sahlberg}
\affiliation{Department of Materials Chemistry, Uppsala University, P.O. Box 538, SE-751 21 Uppsala, Sweden}
\author{E. Br\"{u}ck}
\affiliation{Fundamental Aspects of Materials and Energy, Faculty of Applied Sciences, TUDelft\\ Mekelweg 15, 2629 JB Delft, The Netherlands}
\author{Y. Andersson}
\affiliation{Department of Materials Chemistry, Uppsala University, P.O. Box 538, SE-751 21 Uppsala, Sweden}
\author{P. Nordblad}
\affiliation{Department of Engineering Sciences, Uppsala University, Box 534, SE-751 21 Uppsala, Sweden}
%\thanks{}


\date{\today}

\begin{abstract}
Magnetic and magnetocaloric properties of high-purity, giant magnetocaloric polycrystalline and single-crystalline \BPChem{Fe\_{2}P} are investigated. \BPChem{Fe\_{2}P} displays a moderate magnetic entropy change which spans over 70 K and the presence of strong magnetization anisotropy proves this system is not fully itinerant but displays a mix of itinerant and localized magnetism. The properties of pure \BPChem{Fe\_{2}P} are compared to those of giant magnetocaloric \BPChem{(Fe,Mn)\_2(P,A)} compounds helping understand the exceptional characteristics shown by the latter which are so promising for heat pump and energy conversion applications.
\end{abstract}
\pacs{75.30.Cr,75.30.Sg,75.50.Cc,75.80.+q,65.40.De}
%\keywords{}

%\renewcommand\thefootnote{\fnsymbol{footnote}}
%\nofootinbib
\maketitle

\section{\label{intro}Introduction}
Recently, magnetic refrigeration based on the magnetocaloric effect has been regarded as a more efficient and environmentally friendly alternative to gas compression-based refrigeration. Amongst the most promising working materials for magnetic refrigeration are those based on \BPChem{Fe\_{2}P} such as \BPChem{(Fe,Mn)\_2(P,A)} where \BPChem{A = As, Ge, Si}\cite{tegus_transition-metal-based_2002, trung_tunable_2009, Dung-Mixed}.

The promise to magnetic refrigeration these materials show lies in the combination of the properties they retain from the parent compound with the easy tailoring of its magnetic properties due to stoichiometric changes. The former,  a first order magnetoelastic phase transition, ensures high magnetic entropy and adiabatic temperature changes while the latter guarantees good working materials over a large temperature span.
However, as well characterized as the  \BPChem{(Fe,Mn)\_2(P,A)} compounds have been in the past decade\cite{tegus_transition-metal-based_2002, trung_tunable_2009, Dung-Mixed}, the magnetocaloric properties of pure  \BPChem{Fe\_{2}P} have received very little attention. 

\BPChem{Fe\_{2}P} crystallizes in the so-called \BPChem{Fe\_{2}P}-type structure (space group $P\bar{6}2m$) where two chemical elements occupy four different crystallographic sites. In the hexagonal structure, Fe occupies two different metal sites, the tetragonal (\BPChem{Fe\_{I}}) 3f site, and the pyramidal (\BPChem{Fe\_{II}}) 3g site, while P occupies the two dissimilar sites, 2c and 1b\cite{rundqvist_structures_1959}. Such distribution of Fe and P atoms in the crystal lattice creates two magnetic sublattices with different interactions as well as magnetic moments:  \BPChem{Fe\_{I}} and  \BPChem{Fe\_{II}} being the low and high moment sites, respectively, with a total moment of \BPChem{$\sim$ 2.9 $\mu$\_{B}/f.u.}\cite{Scheerlinck1978181, tobola_magnetism_1996}. Below its Curie temperature (\BPChem{T\_{C}}), around 214 K, the moments are aligned in a ferromagnetic arrangement along the c-axis\cite{fujii_magnetic_1977}. At \BPChem{T\_{C}} a first order magnetic phase transition to the paramagnetic state occurs.

Early works about \BPChem{Fe\_{2}P} strongly disagree on its \BPChem{T\_{C}} and saturation magnetization, and even the nature of the transition was not clear\cite{lundgren_first_1978,fujii_magnetic_1977} since the only example known at the time of a ferromagnetic to paramagnetic first order magnetic phase transition was that of MnAs, described by Bean and Rodbell\cite{bean_magnetic_1962}. By determining the purity of their samples prior to further characterization, Lundgren et al.\cite{lundgren_first_1978} made it clear that the properties of \BPChem{Fe\_{2}P} are very sensitive to stoichiometric deviations, explaining the differences in saturation moment values. The spread in \BPChem{T\_{C}} is two fold: it is stoichiometry dependent and extremely sensitive to the applied magnetic field\cite{fujii_magnetic_1977}. The first order nature of the transition was first proposed by W\"{a}ppling et al.\cite{waeppling_first_1975} due to magnetoelastic effects observed in M\"{o}ssbauer measurements. However it was only after careful measurements that thermal hysteresis\cite{lundgren_first_1978}, the discontinuity of the lattice parameters\cite{lundgren_first_1978,fujii_magnetic_1977} and a considerable latent heat contribution\cite{beckman_specific_1982} at \BPChem{T\_{C}} were observed, determining once and for all the first order character of the phase transition in question.

Therefore, \BPChem{Fe\_{2}P} undergoes a first order magnetoelastic phase transition which is accompanied by a discontinuity in lattice parameters and a small decrease in volume of about 0.04\% (on heating), but no change in crystal symmetry.

It has been recently suggested from first principles calculations that the origin of the metamagnetic transition in \BPChem{Fe\_{2}P} lies in the nature of the \BPChem{Fe\_{I}} sublattice and its interaction with the \BPChem{Fe\_{II}} sublattice. Below \BPChem{T\_{C}}, while the intra layer interactions in the \BPChem{Fe\_{II}} lattice are strongly ferromagnetic, the \BPChem{Fe\_{I}} lattice is essentially non-magnetic and only acquires moment due to the exchange field generated by the ordering of the \BPChem{Fe\_{II}} sublattice\cite{delczeg-czirjak_ab_2010}. Thus, the \BPChem{Fe\_{I}} sublattice provides a strong coupling to the crystal lattice while the interaction between the \BPChem{Fe\_{I}} and \BPChem{Fe\_{II}} sublattices generates a large magnetization jump. This result has been independently obtained for \BPChem{Fe\_{2}P}-based compounds\cite{Dung-Mixed} as well.

In this work we have characterized high purity poly- and single-crystalline  \BPChem{Fe\_{2}P} not only as a magnetocaloric material in itself but also to better understand the outstanding properties shown by  \BPChem{Fe\_{2}P}-based compounds. As \BPChem{Fe\_{2}P} presents high magnetic anisotropy, we emphasize that the  anisotropic character of the magnetic response needs to be taken into account for the correct determination of the magnetocaloric effects. 

\section{\label{exp}Experimental Techniques}
The polycrystalline sample studied was prepared using the drop synthesis technique\cite{carlsson_determination_1973}. The single crystalline needle used in this study was prepared using the tin-melt technique\cite{Jolibois,zemni_synthesis_1986}. Since the mass of the needle measured is below the precision of most balances, its mass was determined by estimating its volume under an optical microscope and calculating it from the known density of this compound. In this manner also the aspect ratio of the crystal was determined to be about 1/15 with the long direction being the crystallographic c-axis. For the magnetic measurements the single crystalline needle was fixed to a silicon slab for easy handling.

The crystallographic properties of both samples were investigated using X-ray diffraction analysis. For the polycrystalline sample the lattice parameters obtained at 296 K using \BPChem{Cu K$\alpha$1} (\BPChem{$\lambda$} = 1.540598 {\AA}) radiation are a = 5.8661(2) {\AA}  and c = 3.4585(3) {\AA}  and were refined using the software \textsc{unitcell}\cite{holland_unit_1997}.

For the needle, X-ray single crystal diffraction intensities were recorded at 100 K on a Bruker diffractometer equipped with an APEX2 CCD detector and a graphite monochromator. The used radiation was \BPChem{Mo K$\alpha$} (\BPChem{$\lambda$} = 0.71073 {\AA}), and the diffractometer was operated at 50 kV, 40 mA. The initial data collection and reduction was performed using the Bruker APEX software. Crystal structure refinements were performed using the software \textsc{JANA2006}\cite{jana}. The composition was refined to be \BPChem{Fe\_{1.995(2)}P}, where only the \BPChem{Fe\_{II}} site (0.592290 0.000000 0.000000) was fully occupied and the lattice parameters obtained were a = 5.8955(0) {\AA} and c = 3.4493(0) {\AA}.

Comparing the lattice parameters obtained in this work for polycrystalline \BPChem{Fe\_{2}P} and those from the work by Carlsson et al.\cite{carlsson_determination_1973}, it can be concluded that, within error, both the single- and poly-crystalline samples have the same chemical composition. This is further supported by the Curie temperatures of both samples which differ by only one kelvin (see Fig. \ref{MxT-bulk}, \ref{MxT-C} and \ref{TcxH}).

The magnetic measurements were performed in Quantum Design's MPMS5XL, MPMS7 and PPMS9 systems. The magnetic entropy change was calculated from the isothermal data using Maxwell's relation. Notice that Maxwell's relations are, in principle, only valid for second order phase transitions. However, they can be used as a good approximation for first order phase transitions if the magnetization change with temperature and/or field is sufficiently smooth.

\section{\label{results}Results}
\begin{figure}
\vspace{-5pt}
\includegraphics[scale=0.85]{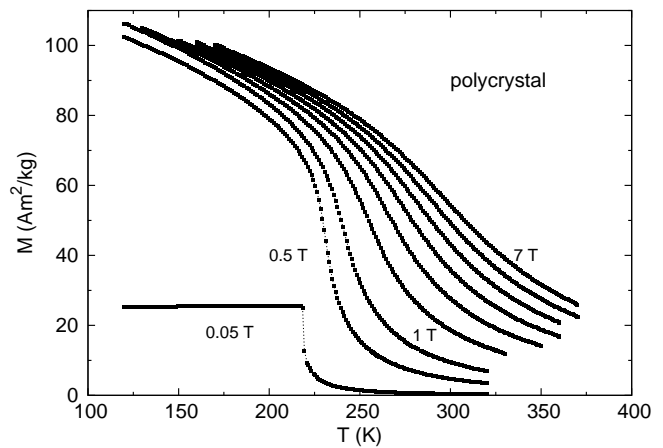}
\vspace{-15pt}
\caption{\label{MxT-bulk}Temperature dependence of magnetization at different magnetic fields for polycrystalline \BPChem{Fe\_{2}P}. The magnetic field intervals between 1 T and 7 T measurements are of 1T.}
\vspace{-5pt}
\end{figure}

\begin{figure}
\vspace{-5pt}
\includegraphics[scale=0.85]{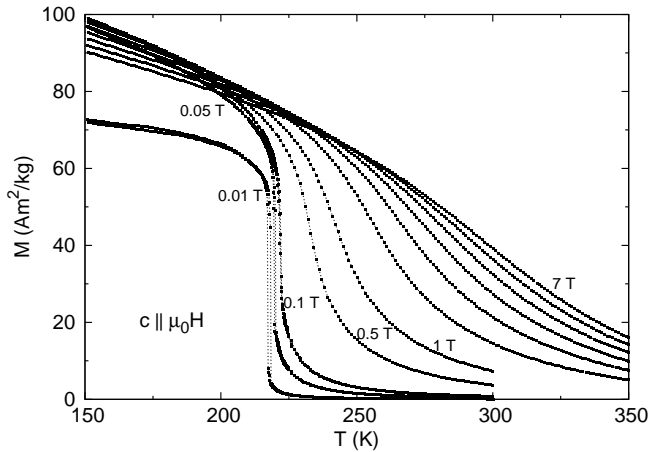}
\vspace{-15pt}
\caption{\label{MxT-C}Temperature dependence of magnetization at different magnetic fields for single crystalline \BPChem{Fe\_{2}P} with the applied field parallel to the c direction. The magnetic field intervals between 1 T and 7 T measurements are of 1T. [Note that in this measurement the magnetization at low temperatures for fields above 1 T is actually lower than for lower fields (see figure \ref{MxT-C}). This is due to the diamagnetic contribution to the magnetization arising from the Si slab where the single crystalline needle was mounted. In the absence of the Si slab the expected behavior would be very similar to that observed in polycrystalline \BPChem{Fe\_{2}P}.]}
\vspace{-5pt}
\end{figure}

\begin{figure}
\vspace{-5pt}
\includegraphics[scale=0.85, clip=true, trim=0 0 0 5pt]{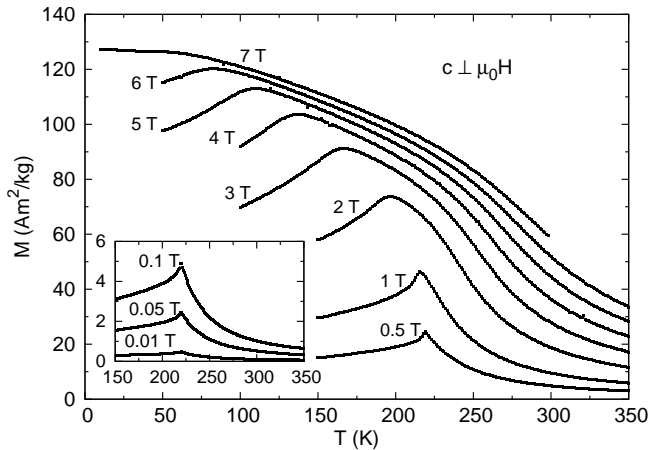}
\vspace{-15pt}
\caption{\label{MxT-Adir}The component parallel to the applied magnetic field of the temperature dependence of magnetization at different magnetic fields for single crystalline \BPChem{Fe\_{2}P} with the applied field perpendicular to the c direction.}
\vspace{-5pt}
\end{figure}

The temperature dependent magnetic properties of polycrystalline and single crystalline \BPChem{Fe\_{2}P} are shown in figures \ref{MxT-bulk} to \ref{MxT-Adir}. For both polycrystal and single crystal in the c-direction the characteristic sharp first order phase transition can be observed for low fields. At \BPChem{$\mu$\_{0}H = 0.01 T} the transition presents a small thermal hysteresis of about 1 K that quickly decreases with increasing field and can no longer be observed for \BPChem{$\mu$\_{0}H $\geq$ 0.1 T}. With increasing magnetic field the transition quickly shifts to higher temperatures and broadens, assuming the characteristics which at first caused \BPChem{Fe\_{2}P} to be considered to have a second order phase transition.

When the single crystalline needle is measured with its hard magnetization axis perpendicular to the magnetic field, i.e. with \BPChem{c $\perp$ $\mu$\_{0}H}, the magnetization direction will be a function of temperature, field and the competition between field driven alignment and magnetocrystalline anisotropy. Thus, the total magnetization \BPChem{M\_{total}} can be represented as a vector which makes an angle \BPChem{$\theta$} with the c axis. The component parallel and perpendicular to the applied magnetic field are given by \BPChem{M\_{$\parallel$} = M\_{total}.$sin\theta$} and \BPChem{M\_{$\perp$} = M\_{total}.$cos\theta$}, respectively, so that the total magnetization is given by \BPChem{M\_{total}\^{2} = M\_{$\parallel$}\^{2} + M\_{$\perp$}\^{2}}.

Figure \ref{MxT-Adir} shows the magnetization component parallel to the magnetic field when the crystal is measured with \BPChem{c $\perp$ $\mu$\_{0}H}. \BPChem{M\_{$\parallel$}}  is deliberately left uncorrected for demagnetizing field. The competition between the temperature and exchange driven spin alignment along the c-direction and the alignment promoted by the applied magnetic field can be clearly observed. Above \BPChem{T\_{C}} the magnetic behavior is dominated by the demagnetization factor of the sample, that is the shape anisotropy. Below \BPChem{T\_{C}} both magnetocrystalline and shape anisotropies compete with the field driven alignment. Up to 0.5 T only one feature can be observed in the temperature dependence of magnetization as a peak in magnetization. Above 0.5 T both this peak and a broad change in curvature at temperatures higher than that of the peak are observed.

\begin{figure}
\vspace{-5pt}
\includegraphics[scale=0.85]{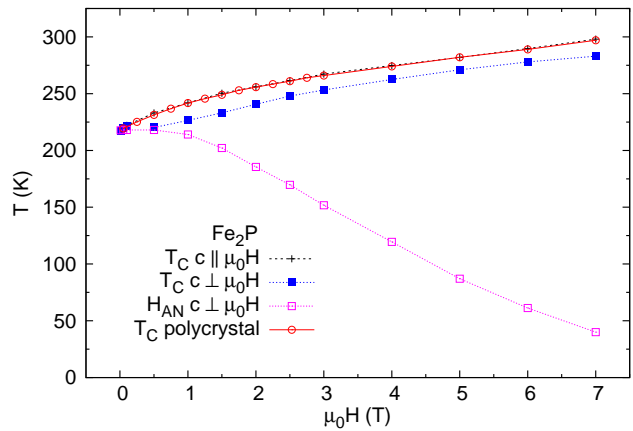}
\vspace{-15pt}
\caption{\label{TcxH}Field dependence of the Curie temperature for several fields for polycrystalline and single crystalline \BPChem{Fe\_{2}P} with the applied field parallel to the a and c directions.}
\vspace{-5pt}
\end{figure}

The field dependence of the critical temperatures for poly- and single-crystalline \BPChem{Fe\_{2}P} is shown in figure \ref{TcxH}. All transition temperatures were taken as the peak observed in the first derivative of the temperature dependence of magnetization. For both polycrystal and single crystal with \BPChem{c $\parallel$ $\mu$\_{0}H} the results are the same. The field dependence of the apparent transition temperature, here referred loosely as \BPChem{T\_{C}}, deviates from a linear behavior for fields below 3 T where it is best fit by a third degree polynomial (\BPChem{ T\_{C} = 217.7(2) K + 30.7(6) K/T $\mu$\_{0}H - 7.9(5) K/T\^{2} ($\mu$\_{0}H)\^{2} + 1.0(1) K/T\^{3} ($\mu$\_{0}H)\^{3}}). Such behavior is in line with previous observations by Fujii et al.\cite{fujii_magnetic_1977} who recorded a shift of 12 K/T for fields below 0.2 T. However, in this work the \BPChem{$\delta$T\_{C}/$\delta$B} observed is much higher than the values obtained by Fujii in the given field interval, reaching approximately 30 K/T. Above 3 T, the increase of \BPChem{T\_{C}} is close to linear with a \BPChem{$\delta$T\_{C}/$\delta$B} value of 7.8(1) K/T.

For measurements performed with \BPChem{c $\perp$ $\mu$\_{0}H} two curves are presented in figure \ref{TcxH}. The temperature at which the peak is observed represents the compensation point of the competition between magnetocrystalline anisotropy and field driven alignment of the spins in the material, which shifts to lower temperatures with increasing field. In other words, it represents the temperature evolution of the anisotropy field and as such is denoted as a field \BPChem{H\_{AN}}. An applied magnetic field of approximately 7 T is necessary to overcome the magnetocrystalline anisotropy at 5 K. The second curve presented is the derivative maximum of the higher temperature broad change in inflection and it follows the trend of \BPChem{T\_{C}} observed when the external magnetic field is applied parallel to the easy magnetization direction, but is shifted about 15 K to lower temperatures. The lower \BPChem{T\_{C}} observed when measuring with \BPChem{c $\perp$ $\mu$\_{0}H} arises from the reduction of the effective field inside the sample caused by the demagnetizing field. This reduction is proportional to the component of the magnetization parallel to the applied magnetic field which is then given by \BPChem{$\mu$\_{0}H' = $\mu$\_{0}H - NM\_{$\parallel$}}, where N = 1/2 is the demagnetization factor when the field is applied perpendicular to a long needle's axis. As a result of the reduction caused by the demagnetizing field, \BPChem{T\_{C}} remains unchanged for \BPChem{$\mu$\_{0}H $\lessapprox$ 0.5 T}, and shows a response equivalent to a lower effective field for higher applied magnetic fields.  

\begin{figure}
\vspace{-5pt}
\includegraphics[scale=0.85]{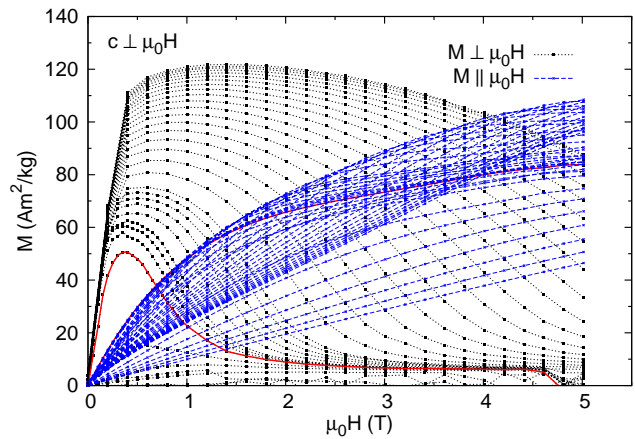}
\vspace{-15pt}
\caption{\label{Fe2P-aIIB-MxHisoT-trans+long} Parallel and perpendicular components of the magnetization as a function of applied field measured with \BPChem{c $\perp$ $\mu$\_{0}H} from 10 K to 360 K. The isotherms were measured using different temperature intervals. Away from the transition, from 10 K to 200 K and from 240 K and 360 K, a 10 K step was used. Closer to the transition region, from 203 K to 212 K and from 221 K to 230 K, 3 k steps were used. Finally, around \BPChem{T\_{C}}, from 215 K to 218 K, the isotherms were measured taking 1 K steps. The isotherm corresponding to \BPChem{T\_{C}} is highlighted in red.}
\vspace{-5pt}
\end{figure}

\begin{figure}
\vspace{-5pt}
\includegraphics[scale=0.85]{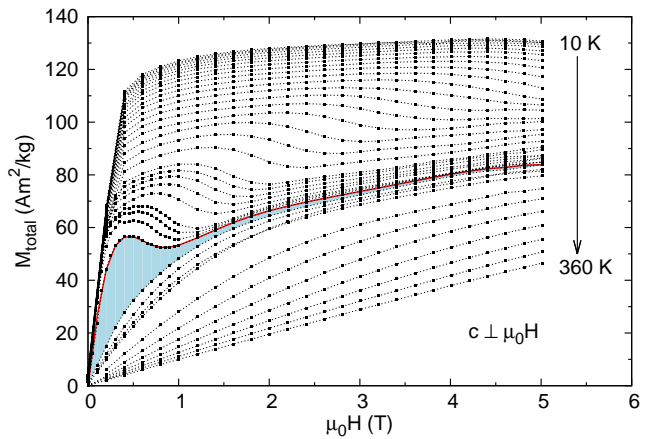}
\vspace{-15pt}
\caption{\label{Fe2P-aIIB-MxHisoT} Total magnetization as a function of the applied field measured with \BPChem{c $\perp$ $\mu$\_{0}H} at different temperatures calculated from the data presented in figure \ref{Fe2P-aIIB-MxHisoT-trans+long}. The large area indicated in blue arises from the interaction between magnetocrystalline anisotropy and field induced transitions at \BPChem{T\_{C}} and the sudden absence of the latter above \BPChem{T\_{C}}.}
\vspace{-5pt}
\end{figure}

\begin{figure}
\vspace{-5pt}
\includegraphics[scale=0.85]{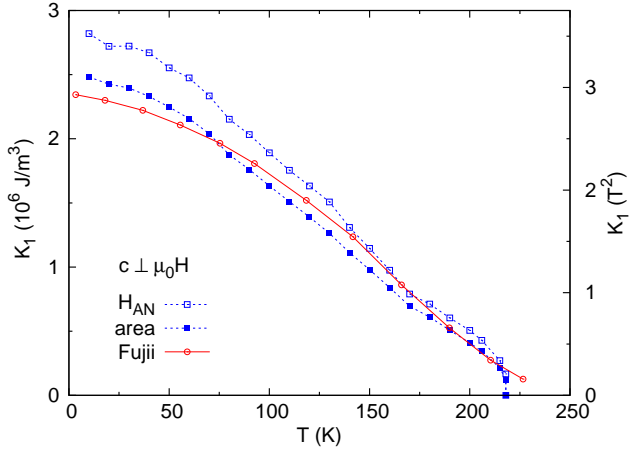}
\vspace{-15pt}
\caption{\label{K1}\BPChem{K\_{1}} calculated from data obtained measuring magnetization parallel and perpendicular to the applied magnetic field while keeping the hard magnetization direction parallel to the latter. Open blue squares represents data calculated using the anisotropy field \BPChem{H\_{AN}}, closed blue squares that from the difference of the areas for isothermal curves and red open circles to the values obtained by Fujii et al.\cite{fujii_magnetic_1977} (red open circles).}
\vspace{-5pt}
\end{figure}

\begin{figure}
\vspace{-5pt}
\includegraphics[scale=0.85]{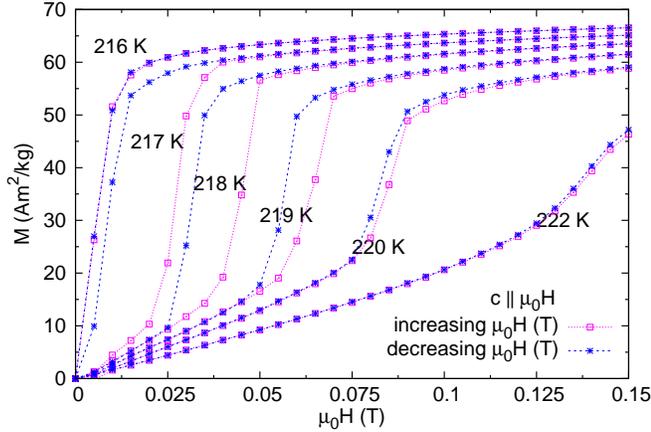}
\vspace{-15pt}
\caption{\label{MxHisoT}Field dependence of the magnetization at several temperatures around \BPChem{T\_{C}} for single crystalline \BPChem{Fe\_{2}P} with the applied field parallel to the c direction.}
\vspace{-5pt}
\end{figure}

\begin{figure}
\vspace{-5pt}
\includegraphics[scale=0.85]{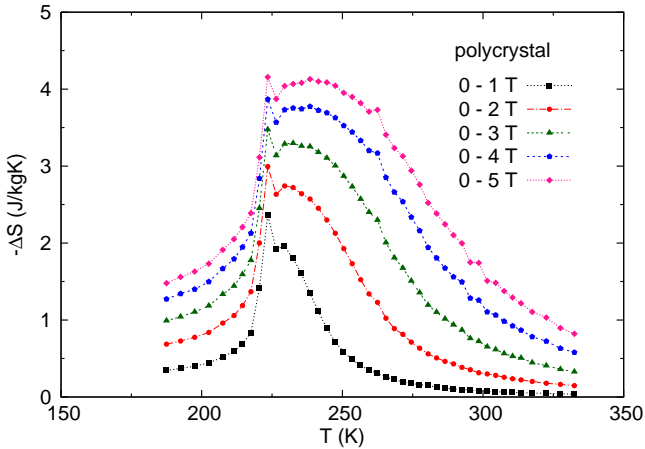}
\vspace{-15pt}
\caption{\label{dSxT-bulk}Magnetic entropy change as a function of temperature for different applied magnetic fields in polycrystalline \BPChem{Fe\_{2}P}.}
\vspace{-5pt}
\end{figure}

\begin{figure}
\vspace{-5pt}
\includegraphics[scale=0.85]{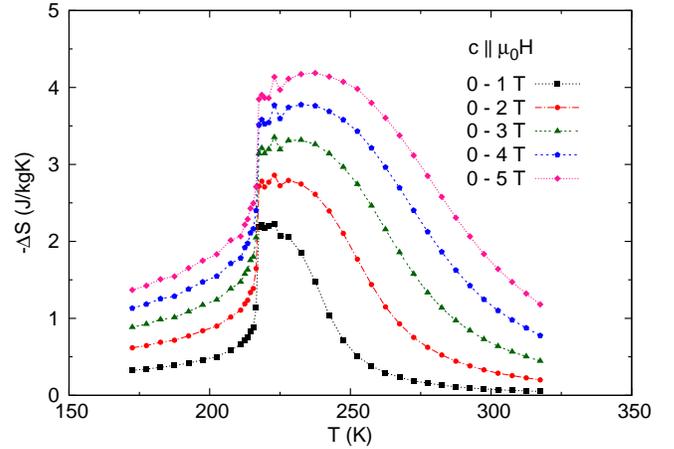}
\vspace{-15pt}
\caption{\label{dSxT-cdir}Magnetic entropy change as a function of temperature for different applied magnetic fields in single crystalline \BPChem{Fe\_{2}P} with the field applied parallel to the easy magnetization direction.}
\vspace{-5pt}
\end{figure}

\begin{figure}
\vspace{-5pt}
\includegraphics[scale=0.85]{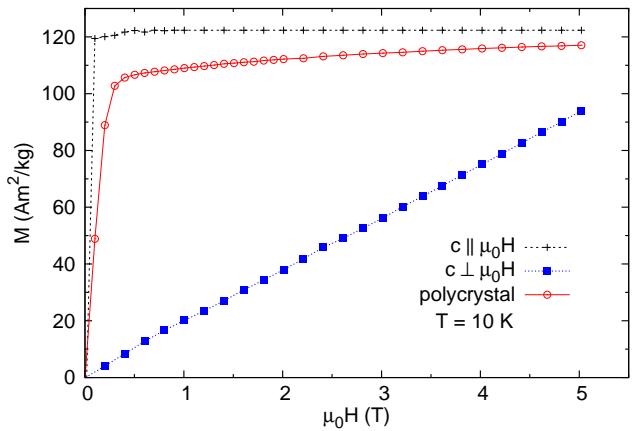}
\vspace{-15pt}
\caption{\label{MxH@10K}Saturation magnetization at 10 K in polycrystalline and single crystalline \BPChem{Fe\_{2}P} with the field applied parallel and perpendicular to the c-direction in the latter case. Notice that the saturation magnetization of the polycrystal is very close to that of the single crystal measured with \BPChem{c $\parallel$ $\mu$\_{0}H}, meaning that our polycrystal is likely to be a collection of well aligned crystallites.}
\end{figure}

\begin{figure}
\vspace{-5pt}
\includegraphics[scale=0.85]{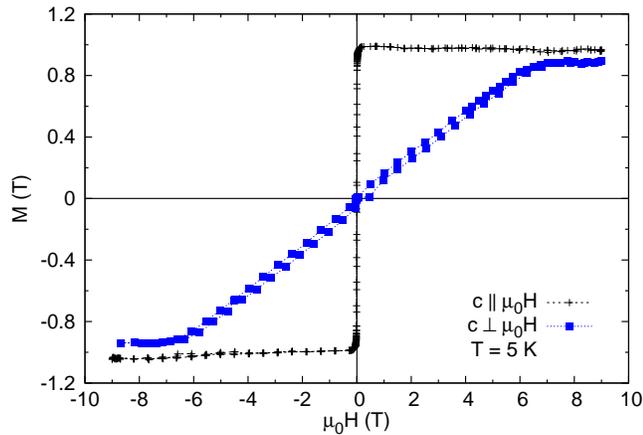}
\vspace{-15pt}
\caption{\label{MxH-Xtal}Field dependence of the magnetization at 5 K in single crystalline \BPChem{Fe\_{2}P} with the field applied parallel and perpendicular to the c-direction.}
\vspace{-5pt}
\end{figure}

The temperature dependence of \BPChem{H\_{AN}} directly reflects the magnetocrystalline anisotropy, which can be more directly evaluated calculating the anisotropy constants. 
A ferromagnetic hexagonal single crystal in the shape of a needle presents, at least, two contributions to the anisotropy energy: magnetocrystalline and shape anisotropies. The magnetocrystalline anisotropy energy is given by: 
\[{E = K_{1}sin^{2}\theta}\]
where \BPChem{K\_{1}} is the first order anisotropy constant and \BPChem{$\theta$} is the polar angle between the c-axis and the magnetization\cite{fujii_magnetic_1977}. This energy is the magnetic work which must be done by the applied magnetic field to bring the magnetization from the easy direction to that imposed by the applied field. This energy can be calculated as the subtraction of the areas under the \BPChem{M\_{T} vs $\mu$\_{0}H} and \BPChem{M\_{$\parallel$} vs $\mu$\_{0}H} magnetization curves (see figures \ref{Fe2P-aIIB-MxHisoT-trans+long} and \ref{Fe2P-aIIB-MxHisoT}) or directly from the extrapolated anisotropy field \BPChem{H\_{AN}}, at a given temperature. Since when \BPChem{M\_{$\parallel$} = M\_{total} $\rightarrow$ $sin\theta$ = 1}, then:
\[W = \int^{\infty}_{0}\left[ M_{total}(H) - M_{\parallel}(H)\right]\mu_{0}dH\] \[W = \frac{1}{2}\mu_{0} H_{AN}M_{total} = K_{1}\]
In figure \ref{K1} the temperature dependence of \BPChem{K\_{1}} calculated using both \BPChem{H\_{AN}} and the difference of the areas are shown. Notice that \BPChem{K\_{1}} calculated using the difference in the areas is slightly underestimated when compared to \BPChem{K\_{1}} calculated from the anisotropy field. The curve obtained by Fujii et al \cite{fujii_magnetic_1977} using the Sucksmith-Thompson\cite{sucksmith_magnetic_1954} method is included for comparison. For the calculation of \BPChem{K\_{1}} and the entropy change when \BPChem{c $\perp$ $\mu$\_{0}H} the applied field was corrected taking into account the shape anisotropy of a needle. All other measurements are presented without corrections.
Isothermal measurements show that poly- and single-crystalline \BPChem{Fe\_{2}P} measured with the magnetic field applied parallel to the easy magnetization direction saturate below 0.1 T. In a very narrow temperature interval around the phase transition a small magnetic hysteresis can be observed and is presented in figure \ref{MxHisoT} for single crystalline \BPChem{Fe\_{2}P} with the magnetic field applied parallel to the c-direction. Notice that a sharp metamagnetic transition can only be observed in the same range where magnetic hysteresis is present.
The magnetic entropy change for both single crystalline and polycrystalline \BPChem{Fe\_{2}P} was calculated from isothermal measurements using the Maxwell relations. As expected, the results for polycrystalline and single crystalline \BPChem{Fe\_{2}P} with \BPChem{c $\parallel$ $\mu$\_{0}H} are very similar (see figures \ref{dSxT-bulk} and \ref{dSxT-cdir}). The magnetic entropy change for the single crystal being slightly higher than that of the polycrystal, due to a higher saturation magnetization presented by the former (see figure \ref{MxH@10K}). The magnitude of the maximum magnetic entropy change for a 1 T field change, 2 and 2.2 J/kgK for poly- and single-crystalline \BPChem{Fe\_{2}P}, respectively, is found to be slightly higher than the 1.8 J/kgK observed by Fruchart et al.\cite{fruchart_magnetocaloric_2005}. Note that the given magnetic entropy values used here for comparison do not take into consideration the sharp peak observed in the low temperature region of the curve, which is found in all measurements presented in both Fruchart's and this work.
\\The use of the Maxwell relation to calculate the entropy change from isothermal measurements in the case where \BPChem{c $\perp$ $\mu$\_{0}H} requires caution. The Maxwell relations are derived from the Gibbs (or Helmholtz)free energy, where the magnetic interaction is included in the form of an energy (or work) which is given by the integral of $\mathbf{M} \cdot \delta\mathbf{H}$, where $\mathbf{H}$ is the effective field and $\mathbf{M}$ the total magnetization. Here the effective field, hereon denoted as $\mathbf{H_{eff}}$, is written as $\mathbf{H_{eff}} = \mathbf{H} + \mathbf{H_{d}} + \mathbf{H_{W}}$, where $\mathbf{H}$ is the applied magnetic field, $\mathbf{H_{d}}$ is the demagnetizing field and $\mathbf{H_{W}}$ is the field due to the exchange interaction with neighboring moments, i.e. the Weiss field\cite{Weiss}. For an isotropic system, or any anisotropic system where the applied magnetic field is parallel to the easy magnetization direction (and to the moment), a variation in the effective field is equivalent to a change in the applied magnetic field once corrections for shape anisotropy are made, since the field due to the exchange interaction with neighboring moments points in the same direction as the applied magnetic field. However, due to the magnetocrystalline anisotropy, this is not true when the applied magnetic field and the easy magnetization direction are no longer parallel{\protect{\footnote{This is best understood when looking at figure \ref{MxH-Xtal}. When a low field is applied perpendicular to c we see at 5 K a very low magnetic response, though the magnetic moments in the needle are aligned along the c axis and thus the sample is fully magnetized.}}}. 

In the magnetization process of a single crystal aligned with its easy axis perpendicular to the applied magnetic field, the moment or field due to the exchange interaction with neighboring moments - and the effective field - is not parallel to the applied magnetic field. In this case, considering that the demagnetizing field is accounted for, a change in the effective field felt by the single crystal results from changes in two components: the applied field and the field due to the exchange interaction with neighboring moments. In order for the magnetization change to reflect a change in both these components it is not enough to consider only the component in the magnetization along the applied field direction, and the total magnetization needs to be considered. In this manner the magnetic entropy change will reflect the change in configurational entropy of the microscopic magnetic moment. Thus, the total magnetization i.e. the magnitude of the magnetization vector, should be used as input of the Maxwell relation. To make our data comparable with the literature our entropy change is calculated with respect to a field change in applied field instead of the effective field.

\begin{figure}
\vspace{-5pt}
\includegraphics[scale=0.85]{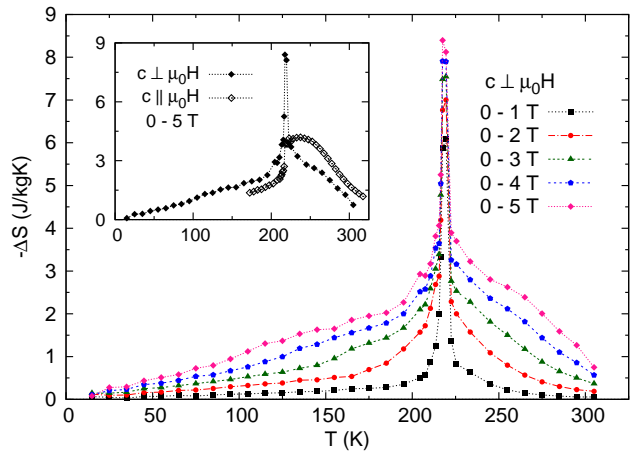}
\vspace{-15pt}
\caption{\label{dSxT-aIIB-Mtotal}Magnetic entropy change as a function of temperature for different applied magnetic fields in single crystalline \BPChem{Fe\_{2}P} with the field applied parallel to the hard magnetization direction. The inset shows the entropy change as a function of temperature for a 0 - 5 T field change measured with the c-axis parallel and perpendicular to the applied magnetic field.}
\vspace{-5pt}
\end{figure}

Therefore, the components of the magnetization parallel and perpendicular to the field (see figure \ref{Fe2P-aIIB-MxHisoT-trans+long}) must be measured and vectorially added resulting in the total magnetization, presented in figure \ref{Fe2P-aIIB-MxHisoT}. The total entropy change calculated from the computed total magnetization is shown in figure \ref{dSxT-aIIB-Mtotal}. Notice that the entropy change curves show a pronounced peak reaching values twice as high as the one observed in the \BPChem{c $\parallel$ $\mu$\_{0}H} case (see figure \ref{dSxT-cdir}). Numerically, this peak is the direct result of the large area observed at low fields in the isothermal data around the first order phase transition, indicated in blue in figure \ref{Fe2P-aIIB-MxHisoT}. In turn, this large area spans from the magnetocrystalline anisotropy and its interaction with the first order phase transition in \BPChem{Fe\_{2}P}.

To understand the reason for this peculiar behavior a more detailed analysis of the magnetization process is required. First we look into the separate components of the magnetization when the crystal is aligned with its easy axis perpendicular to the applied magnetic field. In figure \ref{Fe2P-aIIB-MxHisoT-trans+long} one can see that as the applied magnetic field is increased the component of the magnetization perpendicular to the magnetic field presents an initial increase, due to the alignment of domains. Subsequently, the magnitude of the magnetization response decreases as the moment rotates towards the magnetic field direction, and this decrease becomes sharper as temperature increases and magnetocrystalline anisotropy is reduced.  The counterpart of this can be observed in the component of the magnetization parallel to the magnetic field which increases as the perpendicular component decreases. When the components of the magnetization parallel and perpendicular to the field are added a different scenario than that observed when \BPChem{c $\parallel$ $\mu$\_{0}H} is obtained (see figure \ref{Fe2P-aIIB-MxHisoT}). In all curves above \BPChem{T\_{C}} an initial increase of the magnetization is observed, followed by a small decrease which is then overcome so that the magnetization keeps increasing and saturates. Whereas above \BPChem{T\_{C}} the magnetization increases monotonically. 

The different behaviors below and above \BPChem{T\_{C}} are easily understood considering that the anisotropy field disappears above \BPChem{T\_{C}} as magnetic ordering is lost. However, below \BPChem{T\_{C}} the influence of magnetocrystalline anisotropy can be clearly observed as the slight dip in the magnetization curves which becomes more pronounced around the first order phase transition. As can be observed in figure \ref{MxHisoT} a field induced transition can only be observed at very low fields and at a narrow temperature interval around the first order phase transition when \BPChem{c $\parallel$ $\mu$\_{0}H}. Since the magnetic moments in \BPChem{Fe\_{2}P} are aligned along the crystallographic c-axis it is straightforward to assume that a field induced transition can only be observed along this axis. This is supported by the measured data in the \BPChem{c $\perp$ $\mu$\_{0}H} case, since no field induced transition is observed in the component of the magnetization parallel to the applied field, i.e. along the hard direction. However, the large area present at low field around \BPChem{T\_{C}} observed in the total magnetization measured with \BPChem{c $\perp$ $\mu$\_{0}H} (blue area in figure \ref{Fe2P-aIIB-MxHisoT}) can only be explained considering the interaction of the magnetocrystalline anisotropy and the field induced transition.

Around \BPChem{T\_{C}} as the applied magnetic field is increased a field induced transition develops in the direction perpendicular to the magnetic field resulting in a sharp increase of magnetization. Notice that, because the effective field along the easy direction when \BPChem{c $\perp$ $\mu$\_{0}H} is lower, the field induced phase transition can be observed at apparent fields higher than in the case where \BPChem{c $\parallel$ $\mu$\_{0}H}. However, competing with that increase is the rotation of the moment in the direction of the applied field and the absence of a field induced transition at higher fields, which effectively results in a decrease of the magnetization above a certain applied field. Once \BPChem{T\_{C}} is crossed both field induced transition and magnetocrystalline anisotropy are absent, resulting in a monotonous increase of the magnetization with increasing applied field. Thus the different behaviors below and above \BPChem{T\_{C}} are responsible for the large peak in the entropy change measured in the hard direction. It is worth noticing that the total entropy, i.e. the area under the entropy change vs. temperature curve measured with \BPChem{c $\parallel$ $\mu$\_{0}H} and \BPChem{c $\perp$ $\mu$\_{0}H} are, within error, the same.

For single crystalline \BPChem{Fe\_{2}P} the low temperature magnetization as a function of the applied magnetic field was measured up to 9 T (see figure \ref{MxH-Xtal}). Since a 7 T field is enough to overcome the magnetocrystalline anisotropy at 5 K the turn of the curve from a non-saturating behavior to a fully saturated ferromagnetic behavior can be observed. Surprisingly, \BPChem{Fe\_{2}P} displays strong magnetization anisotropy: the saturation magnetization when the field is applied perpendicular to the c-direction and enough to overcome magnetocrystalline anisotropy is about 9\% below the easy axis saturation magnetization values.
\section{\label{discussion}Discussion}
Substituting Mn on the Fe site and As, Ge or Si on the P site, the crystalline structure and first order magnetoelastic phase transition of pure \BPChem{Fe\_{2}P} are retained. However, tuning the magnetic properties of \BPChem{Fe\_{2}P} is not as simple as substituting similar elements on one of its sites. The substitution or doping on the P site quickly shifts \BPChem{T\_{C}} up, but also leads to the loss of the first order magnetoelastic coupling. Substituting minute amounts of Mn on the Fe site is enough to induce antiferromagnetism and change the crystallographic structure\cite{fujii_magnetic_1982}. 

This reflects the very delicate balance found in the magnetoelastic coupling of \BPChem{Fe\_{2}P}. Thermal and magnetic hystereses are both quite small, and can only be observed at very low fields (see figures \ref{MxT-C} and \ref{MxHisoT}). Moreover, increasing applied magnetic field quickly broadens the first order phase transition and effectively shifts \BPChem{T\_{C}} to higher temperatures. Such behavior suggests that the energy barrier needed to be overcome to go between paramagnetic and ferromagnetic states is quite low. The high \BPChem{$\delta$T\_{C}/$\delta$B} combined with a low magnetic entropy change imply a weak magnetoelastic coupling which is easily affected by an external magnetic field. From the Clausis-Clapeyron equation it is straightforward to conclude that a large \BPChem{$\delta$T\_{C}/$\delta$B} should result in a low entropy change $\Delta$S: 

\[{\Delta S_{total}(T,\Delta H) = -\Delta M \left(\frac{\delta T_{C}}{\delta B}\right)^{-1}}\]
where $\Delta$M is the change in magnetization due to the transition. Consequently, a low adiabatic temperature change \BPChem{$\Delta$T\_{ad}} is also expected, since it is proportional to the entropy change itself. In this sense the behavior of \BPChem{Fe\_{2}P} is very similar to that of the MnCoSi compound reported by Sandeman et al. \cite{sandeman_negative_2006}. MnCoSi shows an even larger sensitivity of the magnetic phase transition to the applied magnetic field, reaching values of -50 K/T. Accordingly, it also displays low entropy changes, even if the metamagnetic transition survives to very high fields, unlike \BPChem{Fe\_{2}P}. It is worth noticing that the high peak in the entropy change for \BPChem{Fe\_{2}P} when measured with  \BPChem{c $\perp$ $\mu$\_{0}H} is directly reflected in the low field \BPChem{$\delta$T\_{C}/$\delta$B}. For fields below 0.5 T, due to magnetocrystalline anisotropy, \BPChem{T\_{C}} remains virtually unchanged at 218 K, resulting in a lower \BPChem{$\delta$T\_{C}/$\delta$B} and a much higher $\Delta$S than in the case where \BPChem{c $\parallel$ $\mu$\_{0}H}.

\begin{figure}
\vspace{-5pt}
\includegraphics[scale=0.85]{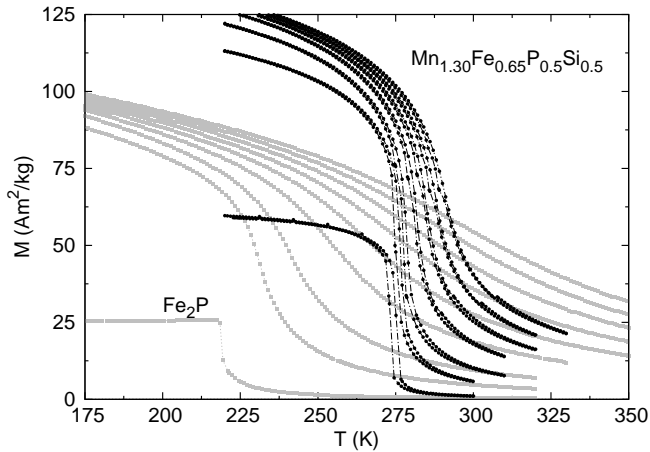}
\vspace{-15pt}
\caption{\label{MxT-comparison}Temperature dependence of the magnetization at different applied magnetic fields for  polycrystalline \BPChem{Fe\_{2}P} and \BPChem{Mn\_{1.30}Fe\_{0.65}P\_{0.5}Si\_{0.5}}. Notice that the magnetic fields used here are the same as in figure \ref{MxT-bulk}.}
\vspace{-5pt}
\end{figure}

\begin{figure}
\vspace{-5pt}
\includegraphics[scale=0.85]{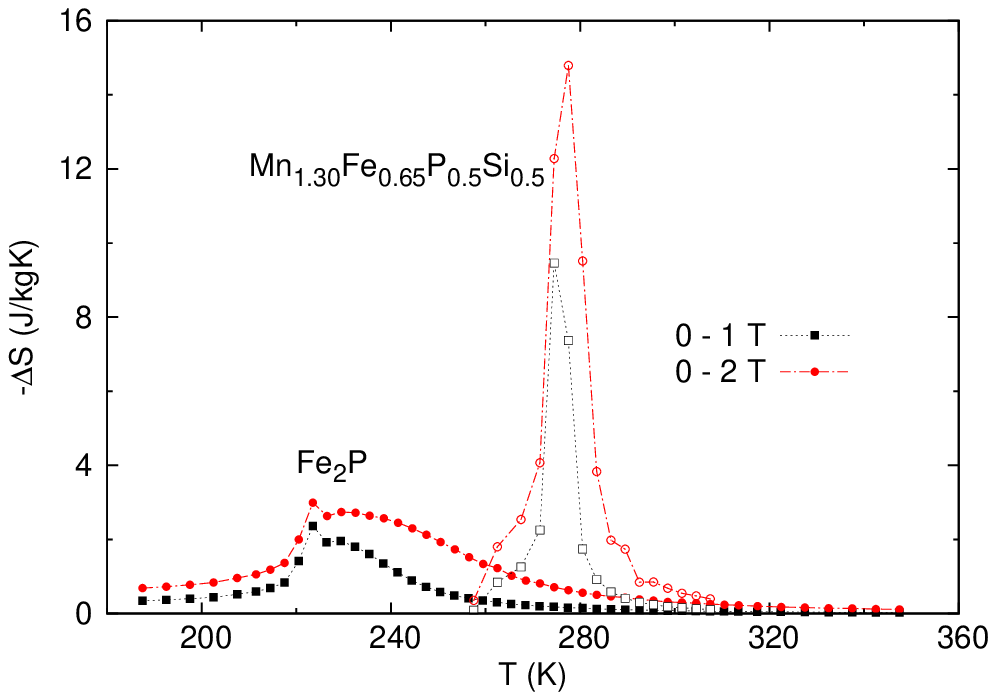}
\vspace{-15pt}
\caption{\label{dSxT-comparison}Temperature dependence of the entropy change for  polycrystalline \BPChem{Fe\_{2}P} and \BPChem{Mn\_{1.30}Fe\_{0.65}P\_{0.5}Si\_{0.5}}.}
\vspace{-5pt}
\end{figure}

These properties are in stark contrast with most \BPChem{(Fe,Mn)\_2(P,A)} compounds, where A = As, Ge or Si. In \BPChem{(Fe,Mn)\_2(P,A)} compounds, while thermal hysteresis can often be reduced by the correct synthesis processing methods, it can always be observed up to very high magnetic fields, around 5 T (see figure \ref{MxT-comparison}). The transition is also hardly broadened by field when compared to pure \BPChem{Fe\_{2}P}. This becomes particularly evident when the field dependence of the Curie temperatures for pure \BPChem{Fe\_{2}P} and \BPChem{(Fe,Mn)\_2(P,A)} materials, \BPChem{$\delta$T\_{C}/$\delta$B} are compared. The first order phase transition in pure \BPChem{Fe\_{2}P} is extremely sensitive to the applied magnetic field, which causes it not only to broaden but also to be shifted to higher temperatures very quickly. In fact the field dependence of the Curie or transition temperature of \BPChem{Fe\_{2}P} is not linear and can be as high as 30 K/T for low fields. In \BPChem{(Fe,Mn)\_2(P,A)} compounds the situation is quite different. The transition is not so easily affected by the applied magnetic field, keeping its first order characteristics up to 5 T and higher. The observed \BPChem{$\delta$T\_{C}/$\delta$B} is linear for \BPChem{(Fe,Mn)\_2(P,A)} materials, as well as comparatively moderate, reaching maximum values of 4 K/T \cite{caron_2009}. Therefore, \BPChem{(Fe,Mn)\_2(P,A)} materials yield much higher \BPChem{$\Delta$S\_{M}} and \BPChem{$\Delta$T\_{ad}} than \BPChem{Fe\_{2}P} (see figure \ref{dSxT-comparison}).
These differences suggest that the energy barrier associated with the first order phase transition in \BPChem{(Fe,Mn)\_2(P,A)} compounds is much higher than in the parent compound. This is also reflected in the size of the lattice parameters change due to the transition in the two cases. The jump in the lattice parameters in \BPChem{(Fe,Mn)\_2(P,A)} compounds is much larger than in \BPChem{Fe\_{2}P}\cite{lundgren_first_1978}. This has intricate consequences which arise from the nature and change of the magnetoelastic coupling in both \BPChem{Fe\_{2}P} and \BPChem{(Fe,Mn)\_2(P,A)} compounds. The key to understanding these materials lies in the coupling of the two different magnetic sublattices. This becomes clear when the interatomic distances are changed. Relatively low pressures are enough to induce antiferromagnetism in \BPChem{Fe\_{2}P}\cite{fujiwara_pressure-induced_1980}. Since the a-direction is the most compressible one\cite{fujiwara_anisotropic_1981}, it is straightforward to assume that pressure decreases \BPChem{Fe\_{I}}-\BPChem{Fe\_{I}} and \BPChem{Fe\_{II}}-\BPChem{Fe\_{II}} more significantly than \BPChem{Fe\_{I}}-\BPChem{Fe\_{II}} interatomic distances. Mn substitution in the Fe site increases the lattice parameters and thus interatomic distances, but since Mn has a higher magnetic moment than Fe the exchange interactions are also larger. Therefore the Mn-Mn interatomic distances are below the critical distance Mn needs to be able to order ferromagnetically\cite{suzuki_magnetic_1973}, resulting in antiferromagnetic ordering instead. Thus the insertion of a larger non-magnetic atom, which acts very much as a spacer, is needed to increase Mn-Mn distances above the critical value where it should order ferromagnetically\cite{liu_temperature_2009}. This is achieved by partially substituting P by As, Ge or Si, enabling not only ferromagnetic order but also recovering the first order character of the transition found in pure \BPChem{Fe\_{2}P}.

Since the \BPChem{Fe\_{I}} sublattice is mainly paramagnetic and acquires moment due to the interaction with the higher moment \BPChem{Fe\_{II}/Mn\_{II}} sublattice, larger lattice parameters mean that a larger change in the phase transition is necessary to bring the system from the paramagnetic to the ferromagnetic state and vice versa. This results in a much larger change in electronic configuration than in pure \BPChem{Fe\_{2}P}, as well as a latent heat contribution at least one order of magnitude higher\cite{beckman_specific_1982,trungthesis}.
The larger magnetic moment of Mn considerably enhances the exchange field generated by the \BPChem{Mn\_{II}/Fe\_{II}} sublattice which in turn causes a much sharper and marked change in the \BPChem{Fe\_{I}} sublattice. This is in agreement with first principle calculation results obtained by Delczeg-Czirjak et al.\cite{delczeg-czirjak_ab_2010} which show that the structural effects in doped and substituted \BPChem{Fe\_{2}P} compounds strengthen the magnetic interactions relative to pure \BPChem{Fe\_{2}P}.

Similar calculations on \BPChem{(Fe,Mn)\_{2}(P,Si)} compounds also point to a stronger magnetoelastic coupling and to a similar interaction between the magnetic sublattices. As in pure \BPChem{Fe\_{2}P} the \BPChem{Mn\_{II}/Fe\_{II}} sublattice generates a large exchange field which induces order in the weakly paramagnetic \BPChem{Fe\_{I}} sublattice\cite{Dung-Mixed}. In terms of the coupling of each magnetic sublattice to the crystal lattice, this means that two distinct behaviors are present. The fact that the \BPChem{Fe\_{I}} sublattice is mostly non-magnetic above \BPChem{T\_{C}} means that the valence electrons contribute to the bond and do not generate moment, having an itinerant character and providing strong coupling to the crystal lattice. The situation is quite different for the \BPChem{Mn\_{II}/Fe\_{II}} sublattice. In the latter, the valence electrons generate high moments which are not lost in the paramagnetic state. This may point at a more localized character, or that a mix of localized and itinerant characters is present in such site.
This essentially means that a previously believed itinerant electron system in fact presents a mix of itinerant and localized magnetisms.

The observation of magnetization anisotropy in \BPChem{Fe\_{2}P} presents the first experimental evidence to support this last assumption. Let us first consider a purely itinerant system. In such a system all the valence electrons should be located in the conduction band and thus be free to move. Therefore, in a single crystal, once magnetocrystalline anisotropy is overcome by the applied magnetic field, it should not matter in which direction (easy or hard magnetization) the field is applied, the response should be the same. However, if some of the electrons are actually localized, a difference should arise depending on which direction the magnetic field is applied. This is exactly the case for \BPChem{Fe\_{2}P} (see figure \ref{MxH-Xtal}). Moreover, the \BPChem{Mn\_{II}/Fe\_{II}} sublattice presents a higher moment  than that of the \BPChem{Fe\_{I}} sublattice, whereas first principle calculations predict the latter to lose its moment above \BPChem{T\_C}. Thus, it is most likely that the localized character lies in the \BPChem{Mn\_{II}/Fe\_{II}} sublattice.

\section{\label{conclusion} Conclusion}
The magnetic and magnetocaloric properties of high-purity poly- and single-crystalline \BPChem{Fe\_{2}P} have been studied. To the authors knowledge this is the first time that a complete magnetocaloric characterization of pure \BPChem{Fe\_{2}P} is carried out. A low but broad entropy change peak as well as a strong field dependence of the first order phase transition are observed. A unique interaction between magnetocrystalline anisotropy and the first order phase transition was observed while measuring single crystalline \BPChem{Fe\_{2}P} with its hard direction parallel to the applied magnetic field, confirming that not only the moments are aligned in the c direction but also that the first order phase transition is tied to the c-axis. 

Comparison with the known properties of \BPChem{(Fe,Mn)\_2(P,A)} compounds provided considerable insight on the nature of the coupling and thus the origin of the magnetocaloric properties of these compounds. Pure \BPChem{Fe\_{2}P} is found to have a weaker magnetoelastic coupling than \BPChem{(Fe,Mn)\_2(P,A)} compounds, clearly visible in stronger first order characteristics such as thermal hysteresis and larger volume changes found in the latter. This is in good agreement with the first principle calculations of Delczeg-Czirjak et al.\cite{delczeg-czirjak_ab_2010} which also conclude that dopings and substitutions strengthen the magnetic interactions. 

Magnetization anisotropy was found to occur in this system, experimentally showing that a previously believed fully itinerant electron system actually displays a mix of localized and itinerant characters. Further analysis strongly suggests that such localized character is probably present in the \BPChem{Mn\_{II}/Fe\_{II}} sublattice. 
\begin{acknowledgments} 
Financial support from the Swedish Energy Agency (STEM) and the Swedish Research Council (VR) is acknowledged. This work is part of an Industrial Partnership Programme IPP I28 of the 'Stichting voor Fundamenteel Onderzoek der Materie (FOM)' which is financially supported by the 'Nederlandse Organisatie voor Wetenschappelijk Onderzoek (NWO)' and co-financed by BASF Future Business. The authors would like to thank Dr. Niels van Dijk from the Delft University of Technology (TUDelft) for valuable suggestions and discussions.
\end{acknowledgments}

\appendix*
\section{MCE and Magnetic Anisotropy}
\begin{figure}[h]
\vspace{-15pt}
\includegraphics[scale=0.4]{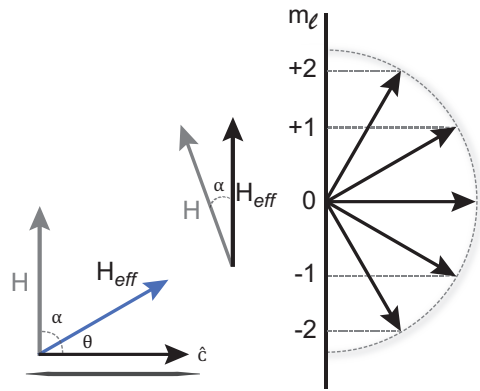}
\vspace{-15pt}
\caption{\label{appendix}[left] Representation of the \BPChem{Fe\_{2}P} needle depicting both applied H and effective \BPChem{H\_{eff}} fields in relation to the c-axis of the needle and the corresponding angles. [right] Vector model of the atom applied to the situation where l = 2 in \BPChem{$\hbar \sqrt{l(l+1)}$} and non zero applied field at an angle $\alpha$ with respect to the effective field.}
\vspace{-5pt}
\end{figure}

An internal magnetic field lifts the degeneracy of the energy levels of the spin (angular momentum) states. This is at the basis of the magnetocaloric effect and from this we immediately also see the applicability of the Maxwell relations, because only the projection of the magnetic moments with respect to the internal field is important to characterize the occupancy of the different energy levels as depicted in figure \ref{appendix}. As described by Weiss and Piccard\cite{Weiss}, the internal field is composed of the applied magnetic field and the field generated by neighboring moments. In a soft ferromagnet these two fields are parallel and we don't need to worry about the moment direction. In a single crystal of a hard magnet this is not the case. Below we give some considerations and experimental evidence for the effect of magnetocrystalline anisotropy.

In the absence of an applied field in the ferromagnetic state the moments are all aligned along the easy axis and due to demagnetizing effects no net magnetic moment is observed. If a perfect crystal is placed with its hard axis exactly parallel to the direction of an absolutely homogeneous applied field, no net magnetization should be observed in the easy axis direction. This can be verified by a simple symmetry argument. However, this ideal situation is hardly ever achieved experimentally, and a moment is always induced along the easy direction. Therefore, to properly evaluate the dynamics of the transition and the magnetocaloric effect of a single crystal under such conditions both components of the magnetization should always be measured, even if it is solely to confirm that your crystal is perfectly aligned!

Here, the fact that we measure a moment in the direction perpendicular to the applied magnetic field when the hard direction is aligned parallel is due to a slight misalignment. Such misalignment can be estimated from the demagnetization factors calculated when measuring the crystal with its easy axis perpendicular and parallel to the applied magnetic field to be around $3\,^{\circ}$. Although this value is within the accuracy of the measurement it also carries the error due to the alignment in two different measurements and therefore must be considered with care. 

This means that the angle $\theta$ is not $90\,^{\circ}$ but $90\,^{\circ} \pm 3\,^{\circ}$. As a consequence a moment is induced along the easy axis causing the total magnetization and the effective field to point at an angle $\theta$ away from the easy magnetization direction or at the complementary angle $\alpha$ away from the applied magnetic field direction. This is represented in figure \ref{appendix}. As expected, with increasing applied field the total magnetization rotates towards the direction perpendicular to the c-axis and parallel to the applied magnetic field. Temperature has a similar effect due to the temperature dependence of the magnetocrystalline anisotropy shown in figure \ref{K1}. 

\begin{figure}
\vspace{-5pt}
\includegraphics[scale=0.85]{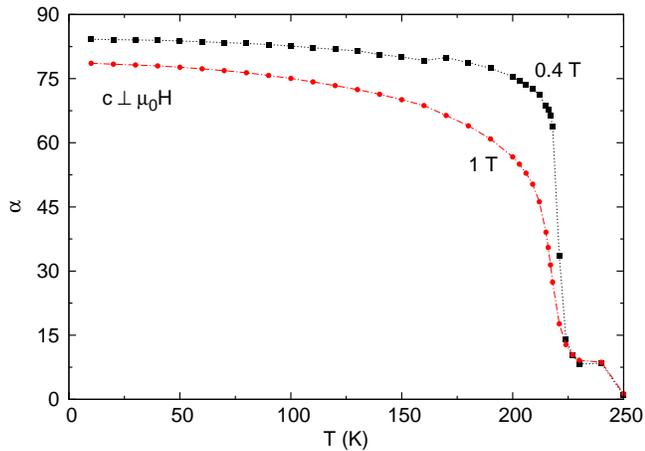}
\vspace{-20pt}
\caption{\label{alpha} Temperature dependence of the angle between the effective field (or  the total magnetization) and the applied field when the latter is applied perpendicular to the easy magnetization direction.}
\vspace{-5pt}
\end{figure}

This can be clearly observed plotting angle isofields as one would do for magnetization. In figure \ref{alpha} the temperature dependence of the angle $\alpha$ at selected applied fields is shown. For low fields (0.4 T) the angle only changes significantly around the first order ferro-paramagnetic transition, at which magnetocrystalline anisotropy disappears. For higher fields the change is more gradual since the magnitude of the magnetic field is comparable to that of the anisotropy field.

\end{document}